\documentclass[apjl,twocolumn]{emulateapj_mod}
\usepackage{epsfig,apjfonts,mathptmx}

\def\gtsima{$\; \buildrel > \over \sim \;$}
\def\ltsima{$\; \buildrel < \over \sim \;$}
\def\prosima{$\; \buildrel \propto \over \sim \;$}
\def\gsim{\lower.5ex\hbox{\gtsima}}
\def\lsim{\lower.5ex\hbox{\ltsima}}
\def\simgt{\lower.5ex\hbox{\gtsima}}
\def\simlt{\lower.5ex\hbox{\ltsima}}
\def\simpr{\lower.5ex\hbox{\prosima}}

\def\h1{$h^{-1}$}
\def\eeq{\end{equation}}
\def\beq{\begin{equation}}

\submitted{Submitted 30 April 2003; Accepted 17 June 2003}

\shorttitle{Near-IR bright galaxies at $z\simeq2$}
\shortauthors{E. Daddi et al.}

\journalinfo{To appear on Astrophysical Journal Letters, special issue for GOODS}
\begin{document}

\title{
Near-IR bright
galaxies at $\lowercase{z\simeq2}$.
Entering the spheroid formation epoch ?$^1$
}

\author{E. Daddi\altaffilmark{2},
	A. Cimatti\altaffilmark{3},
	A. Renzini\altaffilmark{2},
	J. Vernet\altaffilmark{3},
	C. Conselice\altaffilmark{4},
	L. Pozzetti\altaffilmark{5},
	M. Mignoli\altaffilmark{5},
	P. Tozzi\altaffilmark{6},
	T. Broadhurst\altaffilmark{7},
	S. di Serego Alighieri\altaffilmark{3},
	A. Fontana\altaffilmark{8},
	M. Nonino\altaffilmark{6},
	P. Rosati\altaffilmark{2},
	G. Zamorani \altaffilmark{5}
}

\altaffiltext{1}{Based on observations collected at the European
Southern Observatory, Chile (ESO programs 70.A-0140, 168.A-0485), and with the NASA/ESA {\em
Hubble Space Telescope}, obtained at the Space Telescope Science
Institute, which is operated by AURA Inc, under NASA contract NAS
5-26555.}  \altaffiltext{2}{ESO,
Karl-Schwarzschild-Str. 2, D-85748 Garching, Germany
} \altaffiltext{3}{Osservatorio Astrofisico di
Arcetri, L.go E. Fermi 5, Firenze, Italy} \altaffiltext{4}{Caltech, Mail code 105-24, Pasadena (CA) 91125}
\altaffiltext{5}{Osservatorio Astronomico di Bologna, via
Ranzani 1, Bologna, Italy} \altaffiltext{6}{Osservatorio
Astronomico di Trieste, via Tiepolo 11, Trieste, Italy}
\altaffiltext{7}{Racah Institute for Physics, The Hebrew University,
Jerusalem, Israel} \altaffiltext{8}{Oss. Astron. di
Roma, via Dell'Osservatorio 2, Monteporzio, Italy}

\begin{abstract}
Spectroscopic redshifts have been measured for 9 $K$-band luminous
galaxies at $1.7<z<2.3$, selected with $Ks<20$ in the {\em K20 survey} 
region of the Great Observatories Origins Deep Survey area.
Star formation rates (SFRs) of $\sim 100$--500 M$_\odot$ yr$^{-1}$ are
derived when dust extinction is taken into account.  The fitting of
their multi-color spectral energy distributions indicates stellar
masses M$\simgt 10^{11}$ M$_\odot$ for most of the
galaxies. Their rest-frame UV
morphology is highly irregular, suggesting that merging-driven
starbursts are going on in these galaxies. Morphologies tend to be
more compact in the near-IR, a hint for the possible presence of older
stellar populations.  Such galaxies are strongly clustered, with 7 out
of 9 belonging to redshift spikes, which indicates a correlation
length $r_0 \sim9$--17 \h1 Mpc (1 $\sigma$ range).  Current
semianalytical models of galaxy formation appear to underpredict by a
large factor ($\simgt 30$) the number density of such a population of
massive and powerful starburst galaxies at $z\sim2$.  The high masses
and SFRs
together with the strong clustering suggest that at $z\sim 2$ we may
have started to explore the major formation epoch of massive
early-type galaxies.
\end{abstract}

\section{Introduction}

The remarkable success of the Cold Dark Matter (CDM) scenario to account 
for the cosmic microwave background
power spectrum (Bennett et al. 2003), leaves understanding galaxy
formation and evolution as one of the most compelling, unresolved
issues of present cosmology. Semianalytical renditions of CDM hierarchical 
paradigm have so far favored a slow growth with time, with a major fraction 
of the mass assembly taking place at $z\simlt 1$ (e.g., Baugh et al. 2002;
Somerville, Primack, \& Faber 2001), with virtually all massive galaxies 
disappearing by $z\sim 1.5$.
Recent results from the K20 project (Cimatti et al. 2002a,b,c; Daddi
et al. 2002; Pozzetti et al. 2003) appear to be at variance with these 
expectations. The K20 project consists of a spectroscopic survey of 
$\sim500$ $Ks<20$ objects selected over 52 arcmin$^2$, and has revealed
a sizable high redshift tail in the galaxy redshift distribution,
where $\sim 30$ galaxies ($\sim 6\%$ of the total sample) were found
at $z>1.7$. Semianalytical CDM models would have predicted no
galaxy at all at such high redshifts in the whole sample (see Fig. 4
in Cimatti et al.  2002c). The redshift distribution could be
reproduced with pure luminosity evolution (PLE) models, although not for
all realizations (see also Somerville et al. 2003).
However, for only a few among the $z\simgt 1.5$ galaxies was  a
spectroscopic redshift available, 
while for all other such galaxies only the photometric redshifts could be
obtained. In order to put on firmer grounds such major result of the K20 
project (and to understand the nature of these high-$z$ galaxies)
we have conducted new VLT spectroscopic observations of galaxies
with either photometric or (uncertain) spectroscopic redshift above
$z\sim1.7$. 
This letter reports the results of the new spectroscopic
observations, and combines them with the optical HST+ACS and infrared
VLT+ISAAC imaging made available by the Great Observatories Origins Deep Survey 
(GOODS) project (Giavalisco et al. 2003).
We assume a Salpeter IMF;
$\Omega_\Lambda, \Omega_M = 0.7, 0.3$, and
$h = H_0$[km s$^{-1}$ Mpc$^{-1}$]$/100=0.7$.

\section{The spectroscopic sample}
\label{sec:data}
A sample of $20$ galaxies with photometric redshifts $z_{\rm phot}\simgt1.7$ 
were selected among
the 41 galaxies without spectroscopic redshift identification 
in the  32 arcmin$^2$ K20 field that is included in the GOODS-South area.
The photometric redshifts where improved over an
earlier estimate (Cimatti et al. 2002b) by including the ultra-deep
$JHKs$ photometry from the GOODS VLT+ISAAC imaging. 
Within this sample, 10  
objects with most secure $z_{\rm phot}$ have been
observed in November 2002 with VLT+FORS2, 
integrating for 7.2ks with $0''.6$ seeing,
and using the 300V grism covering the range 3600--8000 \AA\ with
13 \AA\ resolution for a $1''$ slit.

\begin{figure}[ht]
\centering 
\includegraphics[width=8.8cm]{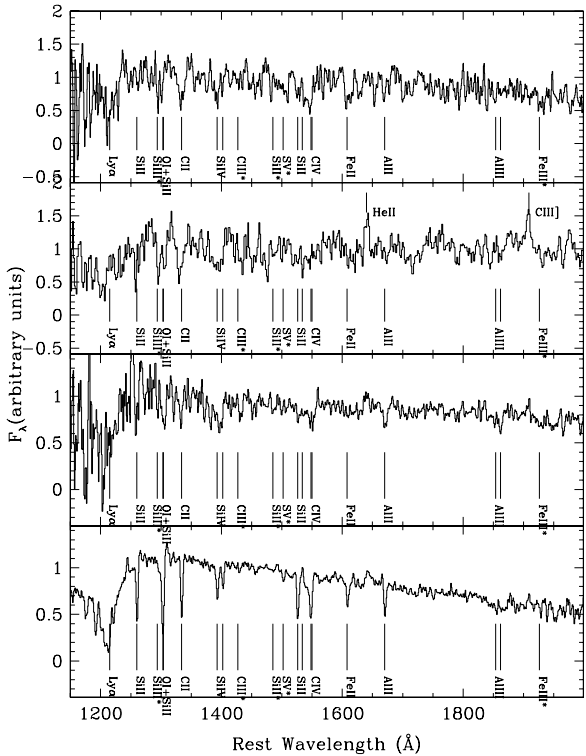}
\caption{The panels show the spectra of, from top to bottom: 
(1) ID\#7 at $z=2.227$; (2) ID\#5, the X-ray and radio source at $z=2.223$;
(3) the average spectrum of 8 $z>1.7$ K-band luminous galaxies 
(ID\#2 has only a near-IR spectrum
showing $H\alpha$ emission)
(4) the average spectrum of $\sim250$ LBGs with Ly$\alpha$ in
absorption (Shapley et al. 2003).
The principal metal lines observed in the UV for starburst
galaxies are marked for reference, namely the ISM lines
SiII$\lambda1260;1304;1527;1534;$, OI$\lambda1303$, CII$\lambda1334.5$,
CIV$\lambda1548;1551$, SiIV$\lambda1394;1403$, FeII$\lambda1608$,
AlII$\lambda1671$,
AlIII$\lambda1855;1863$; as well as the photospheric absorption lines
(shown with an asterisk)
FeIII$\lambda1926$, SV$\lambda1502$, SiIII$\lambda1294$,
SiII$\lambda1485$,CIII$\lambda1427$.
}
\label{fig:spectra}
\end{figure}

The spectra were reduced and calibrated in a standard way (Cimatti et
al.  2003b) and co-added to already existing spectra, when available.
Redshifts were measured for 7 galaxies from absorption features in their blue
continua identified as UV metal lines (Fig.  \ref{fig:spectra}).
For the  $z>2$ galaxies Ly$\alpha$ in absorption and
the onset of Ly$\alpha$ forest are also detected.  
One galaxy (ID\#5) shows HeII$\lambda1640$ and
CIII]$\lambda1909$ emission lines.  Hints for those emission lines are found also
for object ID\#9, for which redshift is less secure because of the
faint and noisy spectrum. Together with 2 previously identified galaxies (ID\#1
and ID\#2), a sample of 9 galaxies with
spectroscopic redshift $z_{\rm spec}>1.7$ is now available among the 304 $Ks<20$
galaxies in the K20/GOODS-South area (Table 1). 
Correspondingly, the fraction of $Ks<20$ 
galaxies with $z_{\rm spec}>1.7$ is $3.0^{+2.8}_{-1.4}$\%,
for a surface density of $0.28^{+0.26}_{-0.13}$
arcmin$^{-2}$, 
and a comoving density of $4.6^{+4.3}_{-2.2}\times 10^{-4}$ 
h$^3$Mpc$^{-3}$
(the range $1.7<z<2.25$ is used, hereinafter, for volume calculations).
The uncertainties are poissonian at the 95\% c.l. 
Accounting for cosmic variance due to clustering 
(Sect. \ref{sec:clustering}) would significantly alter only the upper bounds
(see e.g. Eq.~8 in Daddi et al. 2000).
These densities would increase by a factor of two by 
including the remaining objects with $z_{\rm phot}\simgt1.7$. The good agreement
between $z_{\rm phot}$ and $z_{\rm spec}$ in the poorly tested region $1.7<z<2.3$
(Table 1) suggests a fair fraction of the latter to be genuine $z\sim2$ galaxies.
Obtaining $z_{\rm spec}$ for these is difficult due to their redder colors and thus 
fainter optical magnitudes with respect to galaxies with measured $z_{\rm spec}$.

\section{Properties of K-band luminous galaxies at $z\sim2$}
\label{sec:work}

Using the wealth of subsidiary data available on the GOODS-South area
we investigate in this section the nature of these $K$-band
luminous galaxies with $z_{\rm spec}\sim2$.\\
{\em -- Star formation rates.}
The rest-frame UV spectra of these galaxies indicate that they are
actively star-forming. They are very similar to the template Lyman Break Galaxies
(LBG)
spectrum (Shapley et al. 2003), and often show SiIII$\lambda1294$
absorption due to OB stars (Fig. \ref{fig:spectra}).  Ly$\alpha$
emission is never detected, as for 25--50\% of LBGs (Shapley et al
2003), a hint for significant dust extinction.  The UV fluxes at
2800\AA\ correspond to star formation rates ($SFR$s) in the range of
$\sim10$--40 M$_\odot$yr$^{-1}$ before extinction correction (Kennicutt 1998).  
The extinction at 1600\AA\ was estimated from the UV
spectral slope $\beta$ (Meurer et al. 1999) derived from the
multicolor photometry. 
The $SFR$s derived in this way (adopting the Calzetti et al. 2000 
extinction law) are very high, with a median 
of $\approx 400$ M$_\odot$yr$^{-1}$.
As an alternative estimate {\em hyperz} was used
(Bolzonella et al. 2000) to fit
Bruzual \& Charlot model spectral energy
distributions (SED) to the observed ones from the available deep VLT
BVRIzJHKs photometry, assuming constant $SFR$ (CSF), and solar metallicity.
The best fitting models require reddening in the range 
$E(B-V)\sim0.3$--0.6 and intense
$SFR$s up to $\sim500$ M$_\odot$yr$^{-1}$, with a median of 150 M$_\odot$yr$^{-1}$.
Using a SMC extinction law would lower both the $SFRs$ and $E(B-V)$ estimates.
Interestingly, ID\#5 (which has the  highest extinction-corrected
$SFR$) is a faint (soft) X-ray source (XID563 in the Chandra Deep Field South
catalog, Giacconi et al. 2002). 
If  completely
due to star formation, its X-ray luminosity $L_{\rm 2-10 keV}^{\rm rest}\sim2.7\times10^{42}$ erg s$^{-1}$ cm$^{-2}$ corresponds to 
$SFR\approx 500$ M$_\odot$yr$^{-1}$ (Nandra et al. 2002).
The object is also a faint 1.4~Ghz radio source with a flux density of 
$103\pm13$ $\mu$Jy (Kellermann et al. 2003, private comunication), fully consistent with the
tight X-ray--radio luminosities correlation shown by Ranalli et al. (2003)
for actively star-forming galaxies.
%
We cannot definitively rule out the presence of an obscured AGN, but 
the lack of AGN signatures in its UV spectrum, showing instead 
a strong SiIII$\lambda1294$ photospheric absorption line, the non detection
in the Chandra hard band and the low X-ray to optical flux ratio
($log(f_{\rm 0.5-2 kev}/f_R)\sim -1.3$) indicate this is a vigorous
star-forming galaxy. The stacked X-ray fluxes of the undetected sources 
give a $2\sigma$ detection corresponding to $<SFR>\ \approx 100$ M$_\odot$yr$^{-1}$
($\Gamma=2.1$ is assumed following Brusa et al. 2002). 
In general the limits on the X-ray luminosities
($L_X\simlt 10^{42}$ erg s$^{-1}$ cm$^{-2}$) and the low X-ray to optical 
flux ratios ($log(f_{\rm 0.5-2 kev}/f_R) \simlt -1.5$) imply that our sample 
contains virtually no AGN.
The high $SFR$s qualify these galaxies as {\it starbursts}, and allow to
build up the equivalent of a local $M^*\sim10^{11}$M$_\odot$ galaxy  in 0.2--1 Gyr. 

\begin{figure*}[ht]
\centering 
\includegraphics[width=16.3cm,angle=180]{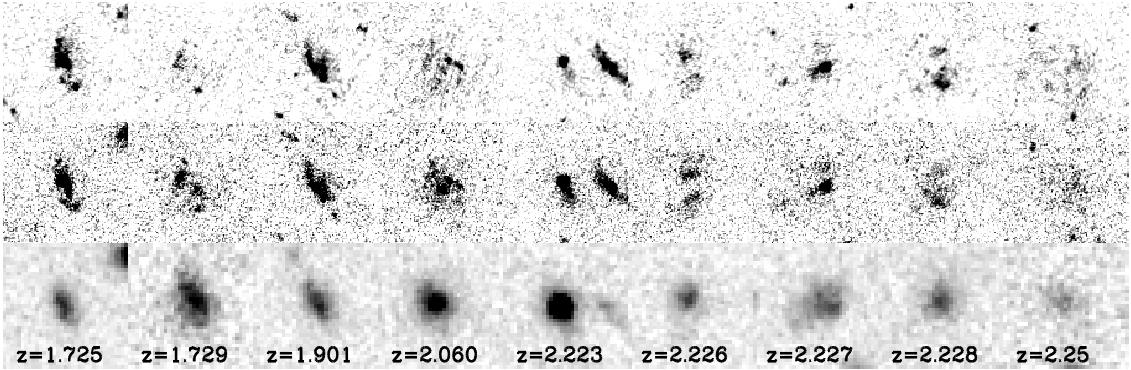}
\caption{ACS (top and center row for F435W and F850LP
respectively, epochs 1+2+3) 
and VLT+ISAAC (bottom row for $Ks$, seeing 0.5$^"$) imaging for the
galaxies with spectroscopic identification. 
The images are 5$''$ on a side. Redshift measurement for each
galaxy is given in the bottom panels.
}
\label{fig:pict}
\end{figure*}

\begin{deluxetable*}{rlllllrcccccccccc}[ht]
\tabletypesize{\scriptsize}
\tablecaption{$K$-band luminous starbursts at $z>1.7$ in the K20/GOODS area}
\tablewidth{0pt}
\tablehead{
\colhead{ID} & \colhead{$z_{\rm spec}$} & \colhead{$z_{\rm phot}$} & \colhead{$Ks$} & \colhead{$J$-$Ks$} & \colhead{$R$-$Ks$}
& \colhead{$\beta$} & \colhead{$SFR_{UV}$} & \colhead{$SFR$} & \colhead{E(B-V)} & \colhead{t} & \colhead{M} & 
\multicolumn{2}{c}{r$_{hl}$} & \colhead{C} & \colhead{A} & \colhead{S}
\\
\colhead{} & \colhead{} & \colhead{} & \colhead{Vega} & \colhead{Vega} & \colhead{Vega}
& \colhead{} & \colhead{M$_\odot$/yr} & \colhead{M$_\odot$/yr} & \colhead{} & \colhead{Gyr} & \colhead{$10^{10}$M$_\odot$}  &
\colhead{$''$} & \colhead{kpc} &  \colhead{} & \colhead{} & \colhead{}
\\
\colhead{} & \colhead{(1)} & \colhead{} & \colhead{} & \colhead{} &
\colhead{} & \colhead{(2)} & \colhead{(3)} & \colhead{(4)} &
\colhead{(4)} &
\colhead{(4)} & \colhead{(4)} & \multicolumn{2}{c}{(5)} &
\colhead{(5)} & \colhead{(5)} & \colhead{(5)} 
}
\startdata
\label{tab:data}
1 & 1.727 & 1.74 & 19.99 & 1.44 & 3.15 & -0.7 & 26 & 93 & 0.3 & 0.36 & 3.4 &  0.6
& 4.8 & 2.2 & 0.21 & 0.9 \\
2 & 1.729 & 2.54 & 19.07 & 1.99 & 4.54 &  0.7 & 13 & 490 & 0.6 & 0.25 & 11 &  0.8
& 7.1 & 1.8 & 0.30 & 1.4 \\
3 & 1.901 & 1.65 & 19.68 & 1.57 & 3.42 & -0.9 & 27 & 155  & 0.3 & 0.52 & 7.9 & 0.6
& 5.4 & 1.9 & 0.41 & 0.6 \\
4 & 2.060 & 1.78 & 19.31 & 1.83 & 4.15 & -0.2 & 35 & 487 & 0.5 & 0.50 & 25 &  0.7
& 5.9 & 2.0 & 0.34 & 1.3 \\
5 & 2.223 & 2.40 & 18.72 & 2.15 & 4.15 &  0.0 & 38 & 540 & 0.4 & 1.0 & 55 &  0.4
& 3.5 & 3.2 & 0.25 & 0.3 \\
6 & 2.226 & 2.24 & 19.94 & 1.69 & 3.30 & -0.4 & 18 & 88 & 0.3 & 0.71 & 6.0 & 0.7
& 5.5 & 2.6 & 0.33 & 0.5 \\
7 & 2.227 & 2.11 & 19.45 & 1.80 & 3.46 & -1.2 & 31 & 178 & 0.3 & 0.72 & 13 & 0.7
& 5.5 & 2.2 & 0.41 & 1.3 \\
8 & 2.228 & 2.43 & 19.74 & 2.02 & 3.66 & -0.4 & 26 & 121 & 0.3 & 1.4 & 17 & 0.9
& 7.8 & 2.1 & 0.25 & 0.9 \\
9 & 2.25 & 2.29 & 19.94 & 1.94 & 3.65 & -0.7 & 30 & 153 & 0.3 & 1.7 & 26 & 1.1 & 9.4 
& 2.7 & 0.29 & 1.2
\enddata
\tablecomments{(1) The redshift for ID\#9 is less secure. (2) UV
spectral slope. Typical errors are $\pm0.1$. (3) $SFR$ derived from the 
2800\AA\ luminosity without
extinction correction. (4) Star formation rate, extinction, luminosity weighted 
stellar age and stellar mass derived from SED fitting of CSF models with reddening.
(5) Quantities measured in the F850LP band, typical errors are 
$\Delta C, \Delta A, \Delta S = 0.15, 0.15, 0.05$.
}
\end{deluxetable*}

{\em -- Stellar masses.}
Although the SEDs are reddened in the rest-frame UV, 
they appear even redder toward the near-IR, where they show 
a steep flux increase, starting in the F850LP band or
beyond, which is suggestive of relatively old stellar populations. 
The CSF models discussed above imply $M_{\rm stars} = 0.3-5.5\times
10^{11}$M$_\odot$, and luminosity-weighted ages of about 250--1700 Myr.
In this case, the near-IR bump is reproduced by a prominent Balmer break.  
The CSF models are likely to underestimate the masses of the galaxies, as
older stars, with higher mass to light ratios, 
may well be present  and yet their light would be outshined 
by the younger ones.  In order to estimate 
reliable upper limits, the minimal
contribution to the $K$-band light by the ongoing starburst
is determined
by fitting a very young ($\simlt 10$ Myr) reddened component to the
SEDs between the $B$ and $I$ bands. This component accounts for 30--50\% of 
the $K$-band light. Assuming the remaining $K$-band light is due
to a maximally old 3 Gyr stellar-population component,
the resulting masses are typically a factor of 2--5
higher than estimated from CSF models, similarly to what is found for 
LBGs (Papovich et al. 2001). 

\label{sec:clustering}
{\em -- Clustering.}
Significant redshift pairing is observed among $z\sim2$ galaxies, 
a clear indication of strong clustering.
Monte~Carlo 
simulations are used to constrain the correlation length $r_0$ from the short
scale pairing, assuming a slope $\gamma=1.8$ for the correlation
function (Daddi et al. 2002).
A flat selection function
between $z=1.7$ and $z=2.25$ is used. 
Seven independent pairs within 5 \h1 comoving Mpc are found in our sample 
of 9 galaxies, implying
$r_0>7$ \h1 Mpc comoving (95\% c.l.), and a most likely range 
 9--17 \h1 Mpc (68\% c.l.).

{\em -- Morphology.}
In the HST+ACS and VLT+ISAAC images taken for the GOODS project
all galaxies show a rather irregular light 
distribution (Fig. \ref{fig:pict}), with bright knots
and low surface brightness regions, often split into separated components.
We measured the {\em CAS} parameters (Conselice 2003; Bershady et al. 2000; and references therein), 
finding relatively high clumpiness ($S$),
high asymmetry ($A$) and very low concentration ($C$) (see 
Table \ref{tab:data}). 
As $S$ is known to correlate with the $SFR$s, the large $S$
values are consistent with the high $SFR$s estimated above.
The $A$ values of most galaxies
are consistent within the errors with the limit of $A>0.35$, typical 
of galaxies undergoing merging or that experienced merging in
the last Gyr (Conselice 2003).  The low $C$ values are
also typical of local merging-driven starbursts, or ultra luminous
infrared galaxies (ULIRGs). There is a trend for increasing 
$C$ from  the rest frame far-UV (F435W band) to the optical 
($K$-band, resolution effects having been taken into account), 
implying a morphological K-correction. Also this is typical 
of starburst galaxies (see e.g. Dey et al. 1999; Smail et al. 2003), 
and may indicate the presence
of an older bulge/disk component (e.g. Labb\'e et al. 2003b) or a higher
reddening in the central regions. All the galaxies appear rather extended,
allowing to host the high estimated $SFR$s.
The average half light radius is $r_{\rm hl}=0''.7$ in the F850LP band, 
about $\sim6$ kpc. 

\section{Relating to other $z\simgt2$ galaxy populations}

We now compare the properties of these $K$-band luminous galaxies to those
of other relevant populations at $z\simgt2$, namely: LBGs at $z\sim 3$,
very red $z>2$ galaxies, and SCUBA sources. 
Compared to LBGs (e.g. Giavalisco et al. 2002), these 
near-IR bright starbursts at $z\sim2$ have, on average,
larger sizes, higher
masses and $SFR$s,  and stronger clustering.
Despite their spectral similarity,
these galaxies  are not just a special subsample of LBGs.
Indeed, it appears that they  have redder UV continuum than the reddest 
LBGs template (Fig. \ref{fig:spectra}). 
In fact, most objects in Table 1 have $\beta>-1$, 
while most LBGs have UV slopes between 
$\beta=-2$ to $-1$, and virtually none has $\beta>-0.5$ (Adelberger \&
Steidel 2000). Hence, the two populations appear only partially overlapping.

These $z\sim2$ starbursts are red in the near-IR, with
$J-K\simgt1.7$, and their clustering is consistent with that of much
fainter $Ks<24$ galaxies at $z>2$ with $J-K>1.7$ colors
(Daddi et al. 2003).  In fact, van Dokkum et al. (2003) 
found significant redshift pairing among 5
galaxies at $z>2$ selected with $J-K>2.3$
(Franx et al. 2003; Labb\`e et al. 2003a). 
Nevertheless, the two samples show different properties,
as strong
$Ly\alpha$ emissions, regular morphologies, and AGN signatures are
common among van Dokkum et al.  objects.
Our 9 spectroscopically confirmed galaxies have $J-K<2.3$ and very clumpy
and asymmetric morphologies.
We conclude that there is a large variety of
properties among $K$-band bright galaxies at $z>2$, that we are just
starting to explore.  

Given the estimated $SFR$s, redshift range and peculiar morphology,
some of our galaxies are potential SCUBA sources (Chapman 2003a,b).  
Red UV SEDs with $\beta>-0.5$ are indeed common among SCUBA sources
(e.g. Dey et al. 1999, Chapman et al. 2002), which also appear to have
large clustering (e.g.  Webb et al. 2003).
Nevertheless, SCUBA sources have much lower spatial density and
are often much fainter than $Ks=20$, while some of our sources may
have too low $SFR$s to be submm-bright. Hence, also in this case
the two populations are likely to overlap only partially.

\section{Discussion}
\label{sec:discussion}

These $K$-band luminous starbursts  provide a substantial
contribution to the cosmic $SFR$ density ($SFRD$) at $z\sim2$:
adding up the $SFR$s
from SED modeling we derive $SFRD\sim0.04$ M$_\odot$ yr$^{-1}$ Mpc$^{-3}$ from the
9 spectroscopically confirmed galaxies alone. This estimate is
certainly affected by incompleteness, and yet it already represents
$\sim 30$--60\% of the $SFRD$ within the range $1.5<z<3$ (see e.g. the
compilation by Nandra et al. 2002). 
These galaxies are also among the most massive
systems detected at $z\sim2$. Six objects in our sample
have $M_{\rm stars}>10^{11}$M$_\odot$ (conservative estimates),
resulting in a number density $\sim10^{-4}$ Mpc$^{-3}$ and a mass 
density of $\sim2\times 10^{7}$M$_\odot$ Mpc$^{-3}$, both $\approx10$~\% of the corresponding local
value (Cole et al. 2001). Integrating over the mass function predicted by the 
Baugh et al.  (2002) model at $z=1.92$, one expects on average only 
0.2 galaxies with $M_{\rm stars}>10^{11}$M$_\odot$
within the explored volume,  and even less
if the semianalytical mass function was properly normalized at $z=0$.
Therefore, these semianalytical models underestimate the number of massive 
galaxies at $z\sim 2$ by about a factor of 30, and possibly much more 
given the incompletenesses of our spectroscopic sample. The assembly
of massive galaxies apparently took place at a significantly larger redshift
(earlier epoch) than predicted by the models (see also Genzel et al. 2003).
On the other hand, these $z\sim2$ galaxies are too actively
star-forming and irregular to be consistent with PLE models with high
redshift of formation.  The agreement between the observed redshift
distribution at $z>1.7$ and the PLE model described in Cimatti et
al. (2002c) is therefore likely to be just chance. 

These $K$-band luminous starbursts are very strongly clustered, suggesting 
they
are hosted in very massive and biased environments, which itself argues
for these objects  being quite massive.
At $z<2$ the only known sources with $r_0\simgt7$ \h1 Mpc are old, passively
evolving EROs (Daddi et al. 2000, 2001) and local massive ellipticals (Norberg
et al. 2002). These $z\sim2$ galaxies are therefore likely to evolve into
such classes of objects.
If star-formation ends rapidly, it would take them $\simgt 1$ Gyr 
to develop very red optical to near-IR colors and to morphologically relax to
regular bulge-dominated galaxies.  This scenario, with massive spheroids 
still forming at $z\sim2$--3, would be quite in good agreement with
some properties of $z\sim1$ old EROs, including
their number counts (Daddi, Cimatti \& Renzini 2000), 
hints for residual star-formation present in their UV
rest-frame (McCarthy et al. 2001), and with the
inferred formation redshifts ($2.4\pm0.3$) of their stellar populations
(Cimatti et al. 2002a). At the same time, it would predict
a paucity of passive EROs at, say, $z>1.3$--1.5.
Finally, we notice that a major shift seems to happen for the clustering
properties of star-forming galaxies from $z\sim1$, where they have very low
clustering
(see e.g. Daddi et al. 2002), to $z\sim2$, where they have a very large
one. The straightforward interpretation is that while
at $z\simlt1$ star formation is mostly confined to low-mass galaxies,
at $z\sim2$ we are starting to see the major build up phase 
of massive early-type galaxies. 
It remains to be determined whether this $z\sim2$ activity represents
the peak or the low-$z$ tail of the massive spheroid formation epoch. 
With the existing technology we should soon be able to answer this question, 
mapping massive galaxy assembly as a function of both redshift and 
large scale structure environment.\\
\acknowledgments
We are in debt to K. Kellermann for measuring the radio flux density
of ID\#5. We thank J. Bergeron, M. Brusa, R. Fosbury, M. Franx and V. Mainieri for
discussions, C. Steidel for providing the LBG composite spectra, and the referee,
J. Primack, for useful comments.

\citeindexfalse

\end{document}